\documentstyle[psfig]{mn}

\title{X-ray Spectroscopy of the IP PQ Gem}

\author[C. H. James, G. Ramsay, M. Cropper
 and G. Branduardi-Raymont]
{Cynthia H. James, Gavin Ramsay, Mark Cropper
and Graziella Branduardi-Raymont\\
Mullard Space Science Laboratory, University College London,
Holmbury St.Mary, Dorking, Surrey, RH5 6NT\\
}

\begin{document}

\maketitle

\begin{abstract} 

Using {\sl RXTE} and {\sl ASCA} data, we investigate the roles played
by occultation and absorption in the X-ray spin pulse profile of the
Intermediate Polar PQ Gem. From the X-ray light curves and
phase-resolved spectroscopy, we find that the intensity variations are
due to a combination of varying degrees of absorption and the
accretion regions rotating behind the visible face of the white dwarf.
These occultation and absorption effects are consistent with those
expected from the accretion structures calculated from optical
polarisation data. We can reproduce the changes in absorber covering
fraction either from geometrical effects, or by considering that the
material in the leading edge of the accretion curtain is more finely
fragmented than in other parts of the curtain. We determine a white
dwarf mass of $\sim$1.2 using the {\sl RXTE} data.

\end{abstract}


\vspace{2.5cm}

\section{Introduction}
\label{intro}

Magnetic cataclysmic variables (MCVs) can be split into two 
groups: those in which the magnetic field of the accreting white dwarf
is strong enough ($\ga$ 10MG) to synchronise its spin period with that
of the binary orbital period -- the polars, and those with a magnetic
field insufficiently strong to achieve this synchronisation -- the
intermediate polars (IPs).  The MCV PQ Gem is unusual in that it
exhibits characteristics of both groups: it shows a strong soft X-ray
component ($kT\sim$50eV) (Duck et al. 1994), it is polarised in the
optical/IR wave-bands (Potter et al. 1997; Piirola, Hakala \& Coyne 1993) 
 and has an estimated
magnetic field strength of $\sim$8--21 MG (V\"{a}th, Chanmugam \& Frank
1996; Potter et al. 1997; Piirola et al. 1993)
 -- all of which are characteristics of polars. On the other hand
it shows a spin period of 833.4 sec (Mason 1997) and an orbital
period of 5.19 hrs (Hellier, Ramseyer \& Jablonski 1994) which are
typical of IPs. PQ Gem can therefore be thought of as the first true
``intermediate'' polar (Rosen, Mittaz \& Hakala 1993).

PQ Gem has been observed using several X-ray satellites ({\sl ROSAT},
{\it Ginga}, {\sl ASCA} \& {\sl RXTE}). The X-ray light curves show a 
prominent modulation on
the spin period, in particular, a pronounced dip in the light curve
which is thought to be due to an accretion stream obscuring the main
emission region on the surface of the white dwarf.  Mason (1997) made a study 
of the then available X-ray data to
determine an accurate ephemeris for PQ Gem based on timings of the
dip. However, a detailed study of the spectral information contained
in the {\sl ASCA} data was not undertaken.

This paper is primarily targeted at reaching a greater 
understanding of the interplay between the emission sites and absorption 
which produces the observed modulation of the X-ray light curves. This is 
achieved through analysis of the  {\sl ASCA} spectral  data with supporting 
evidence from the hard X-ray {\sl RXTE} light curves.  The mass of the 
white dwarf is calculated from the {\sl RXTE} spectral data using the 
stratified accretion column model of Cropper et al. (1999, subsequently CWRK). 
A fuller report is available in James (2001).

\section{Observations and Data Reduction}

\subsection{ASCA}

{\sl ASCA} was launched in 1993 carrying two X-ray CCD cameras (SIS) and two 
imaging gas scintillation proportional counters (GIS), and operated until 2000 
(Tanaka, Inoue \& Holt 1994). The SIS detectors 
covered the energy 
range of 0.4 keV to 10keV with an energy resolution of 2\% at 5.9 keV. The GIS 
detectors had an energy range of 0.7 keV to 10 keV with energy resolution of 
8\% at 5.9 keV. Above 8keV the GIS had a greater effective area than the
SIS.

Details of the observations of PQ Gem are given in Table \ref{tab:ainstru}. The SIS detectors 
were configured to  faint data mode  2-CCD (high bit rate) and 1-CCD (medium 
bit rate) clocking modes and the GIS detectors were configured
 to PH-mode. It is not possible to merge data from the different types of 
detector or from different SIS clocking modes without loss of information.
Hence the end product of the data selection and reduction was several data 
files with effective exposures ranging from 24.7--36.9ksec.

\begin{table}	
\begin{center}\small
%\begin{tabular}{p{15mm}p{8mm}p{10mm}p{8mm}p{16mm}}
\begin{tabular}{lllll}
Date      & Instrument &Bit   & Integration & Average         \\
          &            &rate  & time (s)    & cts/s           \\
\hline
1994/11/04 & SIS0  & high   & 26786  & 0.35 \\
           &       & medium & 36910  & 0.38 \\ 
           & SIS1  & high   & 24732  & 0.29 \\
           &       & medium & 34734  & 0.31 \\ 
           & GIS2  & high   & 27771  & 0.25 \\  
\hline
\end{tabular}
\caption{Details of the {\sl ASCA} observations and data used in 
 the analysis for this paper.} 
\label{tab:ainstru}
\end{center}
\end{table}
  
The data selection and reduction were carried out following recommended
procedures (details in James 2001).  The observation was made early in the 
{\sl ASCA}
mission prior to the onset of the degradation of the CCDs due to radiation
damage (Yaqoob et al. 2000). Hence, the background spectra for SIS were created
using the appropriate blank-sky event lists (available as part of {\sl ASCA}
calibration data) together with the data selection criteria corresponding to
each event list. Background spectra for GIS were created in a similar fashion
using blank-sky event lists. Background light curves were created from a
source-free region of the same observation and subjected to the same screening
criteria as for the source light curves.

\subsection{RXTE}

The Rossi X-ray Timing Explorer ({\sl RXTE}) was launched on 30th
December, 1995.  A review of the available instrumentation and the
scientific results of the first years are given in Swank (1998).

Details of the {\sl RXTE} observations of PQ Gem used in this paper
are given in Table \ref{tab:rinstru}.  Only the top Xenon layer was
selected from the five PCA instruments, so as to improve the
signal-to-noise ratio. The standard {\sl RXTE} procedures were used to
select data (details in James 2001). The background was modelled using
the latest faint background model appropriate for the epoch of the
observation, {\it i.e.} files faintl7\_e03v03.mdl and
faint240\_e03v03.mdl.

\begin{table}	
\begin{center}\small
%\begin{tabular}{p{20mm}p{10mm}p{10mm}p{15mm}}
\begin{tabular}{llll}
Date         & Instrument & Integration & Average       \\
              &           & time (s)    & cts/s         \\
\hline
1997/01/27-30 & PCA        & 51216       & 13.9          \\

\hline
\end{tabular}
\caption{Details of the {\sl RXTE} PCA observations and data used in the 
analysis for this paper.} 
\label{tab:rinstru}
\end{center}
\end{table}

\section{Light Curves}
The quadratic spin ephemeris of Mason (1997) was used throughout the analysis 
of the light curves which were heliocentrically corrected. 

\begin{figure*}
\psfig{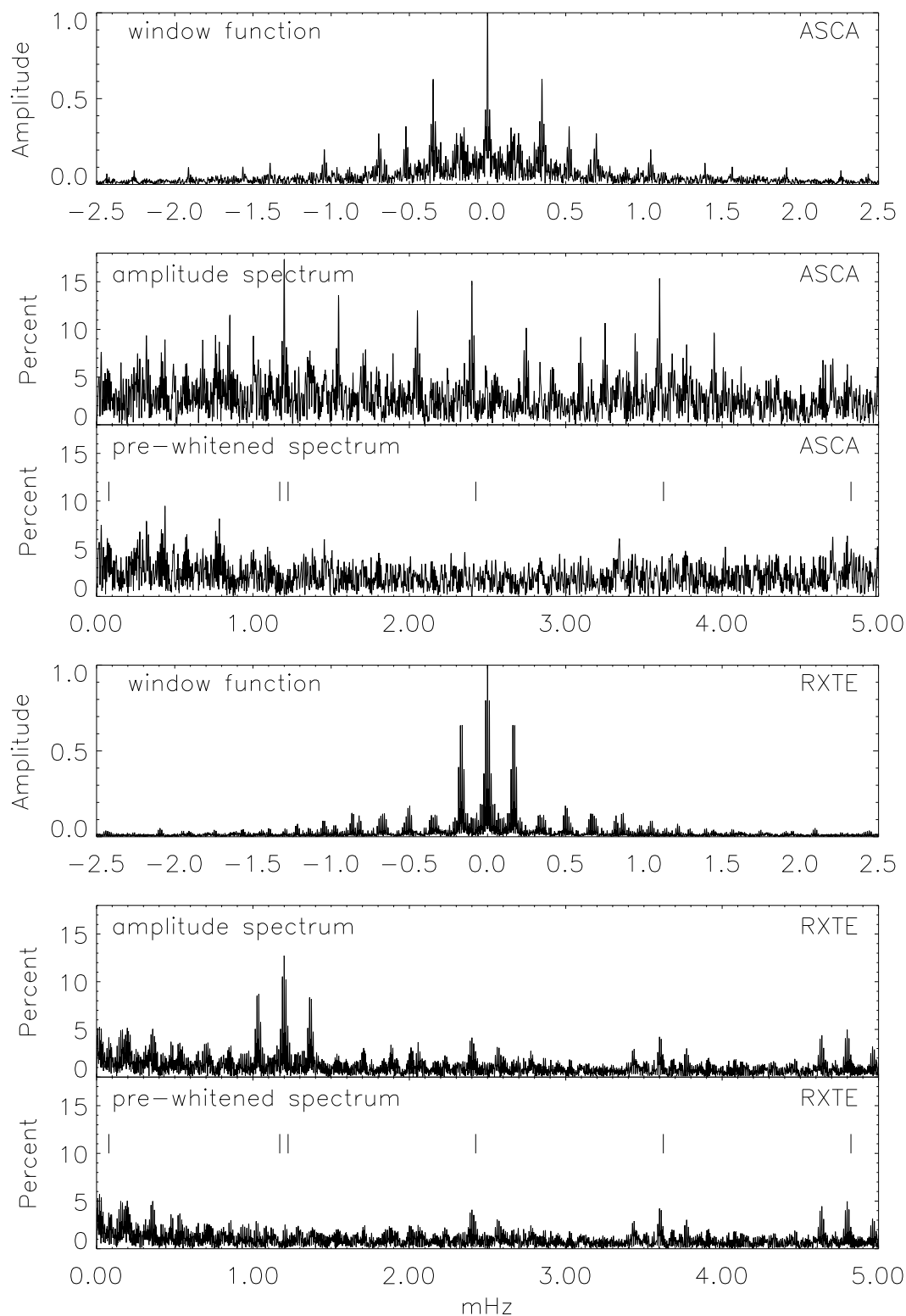}
\caption{Periodograms for the {\sl ASCA} 0.5--10.0keV light curve 
(upper half) and {\sl RXTE} 2.0--25.0keV light curve (lower half) of 
PQ Gem. In each half the top plot shows the window function, in the
middle plot, the 
amplitude spectrum, and the bottom plot shows the spectrum 
prewhitened with the frequencies of the significant peaks. The vertical bars 
(from left to right) mark the position of the orbital, beat and spin 
($\omega$) frequencies  plus the 1st, 2nd and 3rd harmonics of $\omega$, 
respectively. The relevant details are given in \S \ref{period} and Table \ref{tab:period}.} 
\label{fig:period}
\end{figure*}

\begin{figure*}
\begin{minipage}[t]{140mm}
\psfig{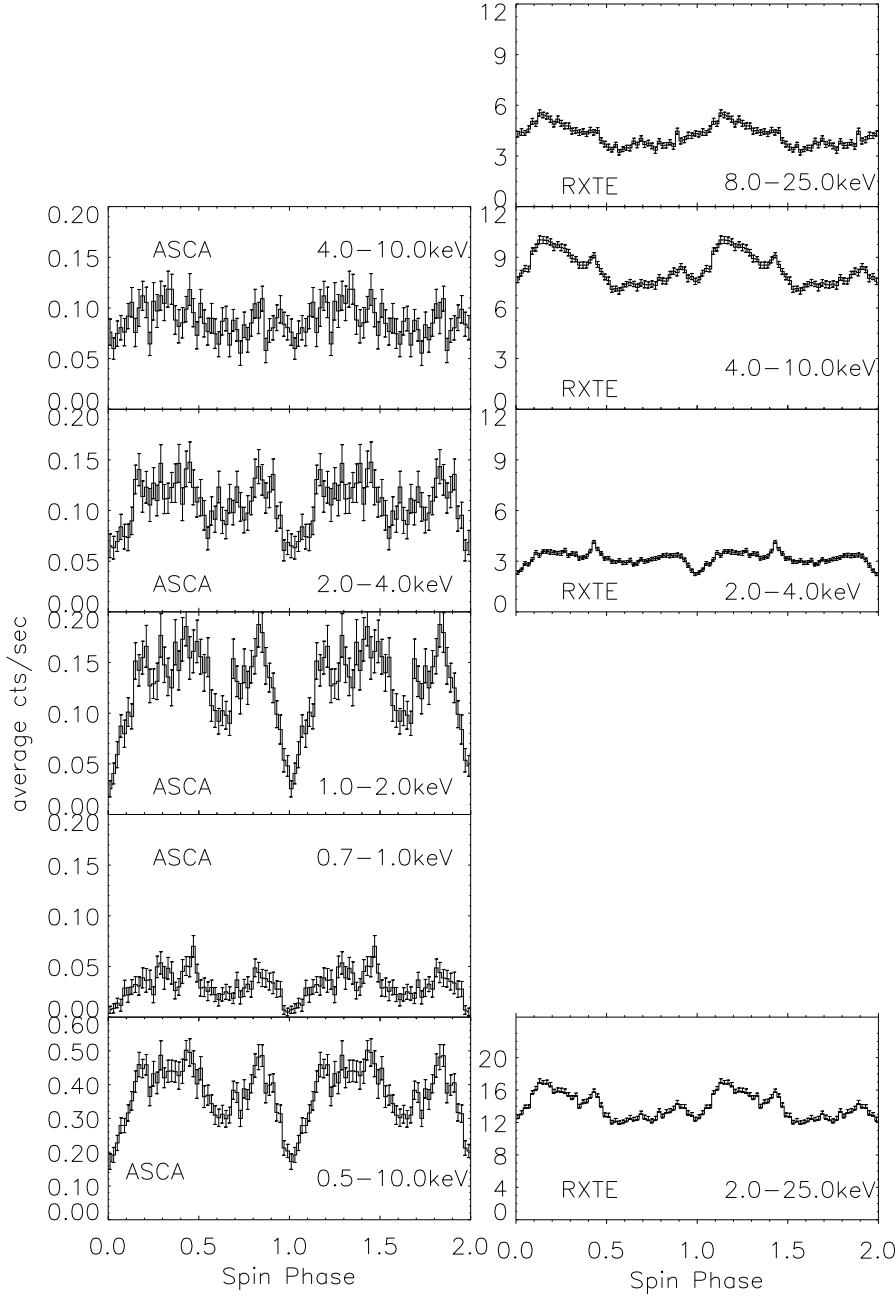}
\caption{Background-subtracted light curves of PQ Gem using {\sl ASCA} SIS0 and {\sl RXTE}
 PCA data folded on the  spin period. On the left
 hand side are the {\sl ASCA} light curves for  energy bandwidths 4.0--10keV (top), 
 2.0--4.0keV (second from top), 1.0--2.0 keV (middle) and 
0.7--1.0keV (second from bottom) 0.5--10.0keV (bottom). On the right hand side
 are the {\sl RXTE} light curves for  energy bandwidths 8.0-25.0keV (top), 
4.0-10.0keV (second to top), 2.0-4.0keV (third to top), 2.0-25keV (bottom).} 
\label{fig:lcs}
\end{minipage}
\hspace*{5mm}
\end{figure*}

\subsection{Period Analysis}
\label{period}

The variability in the {\sl RXTE}  and {\sl ASCA} light curves was 
analysed using a standard discrete Fourier transform (DFT) code (Deeming 1975,
 Kurtz 1985). The equation, $z_0=-ln[1-(1-p_0)^{1/N}]$, (Scargle 1982) was 
used to determine the power level, $z_0$, above which an amplitude peak would 
be spurious for a  small fraction, $p_0$, of the time, where N is the number 
of frequencies examined. A 90\% confidence ($p_0$ = 0.1) was 
used to determine the signal-to-noise level above which the amplitude of a 
power peak was judged to be significant. The noise level of the periodogram 
was taken to be the mean of the amplitude spectrum after it had been 
prewhitened  with frequencies found to be significant (Table \ref{tab:period}). 
The error in each such  frequency was taken to be  the standard deviation, 
$\sigma$, from a least squares fitting.
         
The {\sl ASCA} data showed three significant periods, 833.46s,
416.678s and 277.763s which are consistent with spin ephemeris of
Mason (1997) and its first two harmonics. The {\sl RXTE} data showed a
significant period of 833.54s. The spin period (833 sec) is consistent
with the spin down rate predicted by the ephemeris of Mason (1997).
The complete results are given in Table \ref{tab:period} for those
signals identified as significant at the 90 percent confidence level.
The amplitude spectra, and window functions for the {\sl RXTE} and
{\sl ASCA} are shown in Figure \ref{fig:period}, along with the
spectra prewhitened with the frequencies of the significant power
peaks and the expected positions of the orbital, beat, spin
frequencies and spin harmonics. The residual power peaks in the
prewhitened {\sl RXTE} periodogram are the 3rd harmonic of the spin
period ($>$ 68\% confidence level) and the 1st and 2nd harmonics
($<$68\% confidence level).

Using an orbital period of 5.19hrs (Hellier et al. 1994) a beat period
at 14.54 min may be expected. However, there is no evidence for a
significant amplitude at the frequencies corresponding to the orbital
or beat periods in either the {\sl RXTE} or the {\sl ASCA} data
(Figure \ref{fig:period}). This indicates that the accreting material
must go through a disc since all orbital information will then be lost.

\begin{table*}	
\begin{center}\small
\begin{tabular}{p{15mm}p{15mm}p{20mm}p{30mm}p{30mm}p{20mm}}
Satellite& Instrument& Passband& Frequency & Period      & S/N  \\
         &           & keV     & mHz       & s           &      \\ 
\hline 
{\sl ASCA}& SIS1     &0.7-10.0 &1.19981(25)&833.46(17) &8.0 \\ 
         &           &         &2.39994(32)& 416.678(56) &7.1 \\
         &           &         &3.60019(32)& 277.763(25) &6.6 \\ 
\hline 
{\sl RXTE}& PCA       &2.0-25.0&1.199703(81)& 833.540(56)&12.8 \\
\hline
\end{tabular}
\caption{The details of the significant power peaks found  from period 
analysis of {\sl ASCA} SIS1 data and  {\sl RXTE} PCA data. The signal-to-noise
 ratio was   taken as the amplitude of power peak to the mean of the 
prewhitened spectrum.}
\label{tab:period}
\end{center}
\end{table*}

\subsection{ASCA light curves}

Light curves of these data were first presented in Mason (1997) and also in 
James et al. (1998). They are presented here for completeness and comparison 
with those from {\sl RXTE}. Light curves
using data from the SIS0 detector for energy bands 0.7--1.0keV,
1.0--2.0keV and 2.0--4.0keV, 4.0--10.0keV and from the total bandpass 
(0.7--10.0keV) folded on spin period are shown on the left-hand side of Figure
 \ref{fig:lcs}. The light curves were not folded on either the orbital or beat
 periods due to the lack of amplitude peaks found at these frequencies in the 
power spectrum of the {\sl ASCA} data (\S \ref{period}).     

\subsection{RXTE light curves}

Light curves were extracted for the energy bands 2.0--4.0keV, 
4.0--10.0keV, 8.0--25.0keV and from the total bandpass, 2.0--25keV, which 
were then folded on the spin period (right-hand side, Figure \ref{fig:lcs}).  
The first two energy bands can be directly compared with similar plots using 
the {\sl ASCA} data. 

\subsection{Comparison between the ASCA and RXTE light curves}

 The {\sl RXTE} PCA instrument overlaps the  {\sl ASCA} SIS instruments across
 the 2--10keV energy band. Hence this energy band from both
instruments was used to compare the light curve modulation 
of the two observations (Figure \ref{fig:lcs}).  The  light curves in these 2 
bandpasses were found to be very  similar,  each with two maxima separated by 
a minimum at spin phase 
$\phi$ $\sim0.0$ (referred to as ``the dip'', Mason et al. 1992) and a 
second minimum at $\phi$ $\sim0.6$. In the 4.0--10.0keV light curves the minima
are relatively shallow in both the {\sl ASCA} and the {\sl RXTE} light curves,
 but the 2.0-4.0keV light curves show  well defined ``dips'' at $\phi$ = 0.0, 
with a broader minimum at $\phi$ $\sim0.6$.
There is some evidence that 
the amplitude of the variation in the {\sl RXTE} folded light curve is less 
than that in the  {\sl ASCA} SIS data.  The shape of ``the dip''  in the
 2.0-4.0keV band is ``V''-shaped in the {\sl RXTE} compared to a more 
``U''-shaped in the  {\sl ASCA} light curve and its depth  appears greater 
in the latter. This may be due to a higher level of 
absorption  at the earlier epoch of the  {\sl ASCA} observation.  

\subsection{RXTE hardness ratios}
\label{hardness}

\begin{figure}
\begin{minipage}[t]{80mm}
\psfig{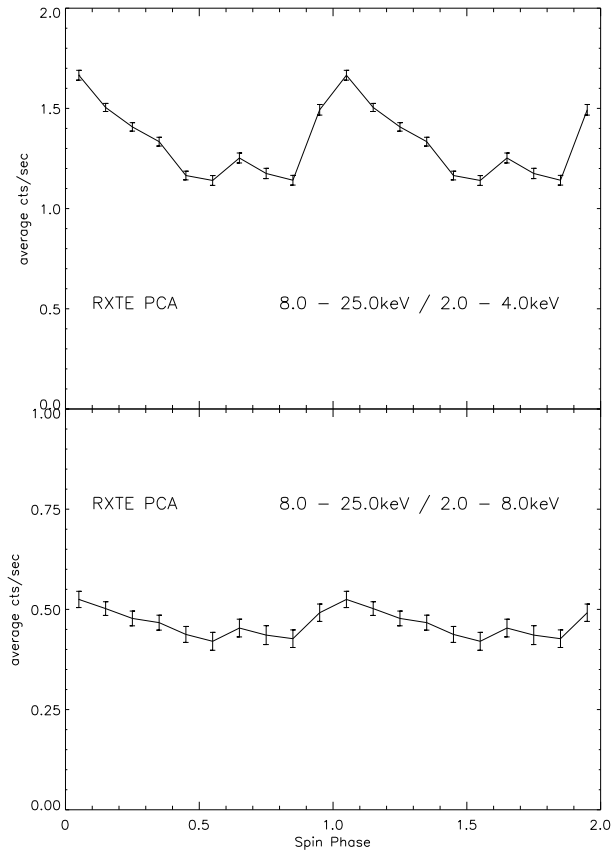}
\caption{Plots of the 8.0--25.0keV/2.0--8.0keV (below) and 
8.0--25.0keV/2.0--4.0keV (above) hardness ratios folded on the 
spin period. The errors are 1$\sigma$.}
\label{fig:hrats}
\end{minipage}
\end{figure}

Plots of the 8.0--25.0keV/2.0--8.0keV and 8.0--25.0keV/2.0--4.0keV
hardness over the spin period are shown in Figure \ref{fig:hrats}. The
lower limit of the 8.0--25.0keV energy band was chosen so as to be
well above the main absorption edges. A higher value of the ratio
indicates those positions in the spin cycle which are more affected by
absorption. These are consistent with the dip at $\phi$ = 0.0 being
due to absorption.

\section{Spectral Analysis}
\label{spctan}
Because the energy resolution of the {\sl RXTE} PCA (18\% at
6.0keV) is much poorer than that of the {\sl ASCA} SIS (2\%), and does
not extend to lower energies, the spectral analysis was restricted to
the {\sl ASCA} data. During the spectral analysis the medium and high
bit rate SIS data from both instruments and the GIS data were linked
through the models and fitted simultaneously.

\subsection{Integrated Spectrum}
\label{intspct}

\begin{figure}
\begin{minipage}[t]{80mm}
\psfig{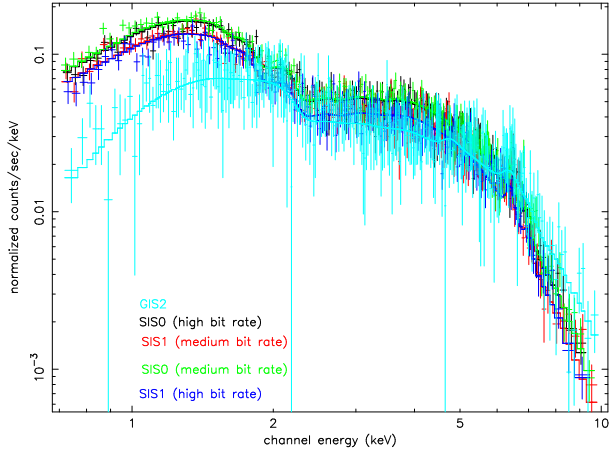}
\caption{Integrated spectra of PQ Gem. Data from the  medium and high bit rate
 data from the SIS1 and SIS0 instruments and the high bit rate data from the 
GIS2 instrument with the fitted model (Table \ref{tab:intspct}).}
\label{fig:intspct}
\end{minipage}
\end{figure}

The integrated spectrum was modelled using the MEKAL code for an
optically thin emission model (Mewe, Kaastra \& Liedahl 1995) for the
hard X-ray spectrum and a blackbody model for the soft X-ray
continuum.  The temperature of the former was difficult to constrain,
a problem which was similarly experienced by Duck et al. (1994) using
{\it Ginga} data. Therefore the temperature was fixed at 20keV in line
with the results from their analysis (we tried fixing the temperature
at other values, but this did not have a significant effect on the
results).  The temperature of the blackbody component was not well
constrained and therefore this spectral component was fixed at 55eV in
line with that found with {\sl ROSAT} data (Duck et al. 1994) due to
its greater sensitivity to this spectral range. The fluorescent iron
line at 6.4keV was modelled using a Gaussian component, the width of
which was fixed at 0.05keV. A single homogeneous photo-electric
absorber gave a poor fit ($\chi_{\nu}^2$ = 2.0), therefore an
inhomogeneous (partial covering) photoelectric absorber was added to
the model. A good fit to the data was achieved in this way
($\chi_{\nu}^2$ = 1.07). The values of those parameters allowed to
vary during the fitting are given in Table \ref{tab:intspct} and the
integrated spectra are plotted in Figure \ref{fig:intspct}.

\begin{table*}	
\begin{center}\small
%\begin{tabular}{p{15mm}p{25mm}p{12mm}p{13mm}p{13mm}p{15mm}p{30mm}p{15mm}}
\begin{tabular}{lllllll}
N$_{H}(1)$              &N$_{H}(2)$ (fraction)                          & BB                & MEKAL                & Gaussian           & Flux(observed)     &$\chi^2_\nu$ (dof) \\
10$^{22}$ cm$^{-2}$    &10$^{22}$ cm$^{-2}$                           & norm:             & norm:                & norm:              & 2-10keV            &                   \\
                       &                                              &
10$^{-3}$         & 10$^{-2}$            & 10$^{-5}$          & 10$^{-11}$erg
cm$^{-2}$ s$^{-1}$     &                   \\
\hline 
0.12$_{-0.03}^{+0.02}$ &8.3$_{-1.3}^{+0.9}$ (0.53$ _{-0.02}^{+0.01}$)&2.0$_{-0.4}^{+2.3}$&1.71$_{-0.06}^{+0.05}$& 4.9$_{-0.9}^{+1.3}$& 2.12               &1.07 (1070)      \\
\hline 
\end{tabular}
\caption{ Results from spectral analysis
 of the {\sl ASCA} integrated spectrum of PQ Gem including the error range at 
90\% confidence level. The optically thin plasma and blackbody temperatures 
were fixed at 20keV and 55eV respectively. N$_{H}(1)$ corresponds to the 
homogeneous absorber whereas N$_{H}(2)$ applies to the inhomogeneous (partial 
covering) absorber; the flux is not corrected for absorption. The model was 
fitted simultaneously to the medium and high bit rate data from the SIS1 and 
SIS0 instruments and the high bit rate data from the GIS2 instrument.} 
\label{tab:intspct}
\end{center}
\end{table*}

\subsection{Phase-resolved Spectroscopy}
\label{phasespect}

Even though a good fit to the integrated spectrum was obtained, the results
from the 8.0--25.0keV/2.0--8.0keV and 8.0--25.0keV/2.0--4.0keV hardness ratios 
using the {\sl RXTE} PCA data indicate the possibility of variation in the 
absorption during the spin cycle
(Figure \ref{fig:hrats}, \S \ref{hardness}).  These two results
imply that variability in both the absorption and the normalisation of the
hard X-rays over the spin cycle could be the cause of the observed modulation 
in the spin-folded light curves. In order to investigate this further we 
divided the spin period  into 10 equal time bins and extracted spectra 
for each using data from both the SIS and GIS2 instruments.  
 During all subsequent spectral fitting the medium and high bit rate data from
 both SIS and the GIS2 instruments were linked through the models and fitted 
simultaneously. 

 The reference model was derived from 0.4 $\leq$ $\phi$ $<$ 0.5, since
this phase had the greatest number of counts. The model resembled that
obtained from the integrated spectrum with the exception of the
blackbody component whose temperature was significantly higher (best
fit=83eV). In this more detailed analysis it was considered to be more
appropriate to fix the temperature of the blackbody component of the
other phase resolved spectra at this value.  The best fit model
comprised of three emission components: an optically thin plasma
emission component with a temperature of 20keV (fixed), a blackbody
component with a temperature of 83eV and a Gaussian at 6.4keV to model
the fluorescent Fe emission line (fixed width of 0.05keV). Two
photoelectric absorption components were required to model the
attenuation of the X-rays: an inhomogeneous (partial covering) neutral 
absorber and a
homogeneous neutral absorber. The latter had a best fit of
$6.8\times10^{20}$ cm$^{-2}$: this component was fixed at this value
in the other phase intervals. The metal abundance had a best fit of
0.34 solar - again this was fixed at other phase intervals.
We now allowed the normalisation of the X-ray emission and the partial
covering absorption component to vary during the remaining 9 phase
bins of the {\sl ASCA} data.  We also allowed the normalisation of the
blackbody and the Fe fluorescent emission line to vary.

\begin{figure}
\begin{minipage}[t]{80mm}
\psfig{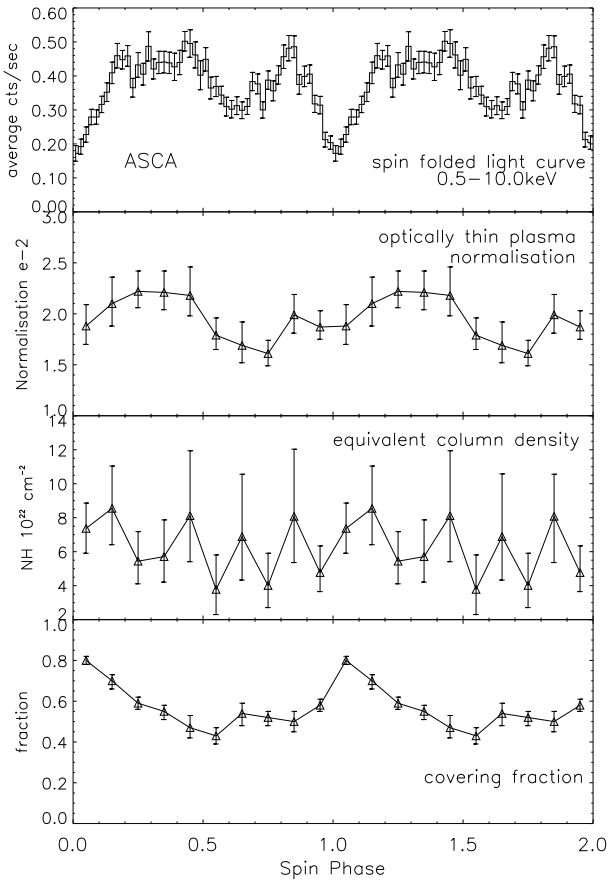}
\caption{Results from spin phased-resolved 
spectroscopy of {\sl ASCA} data. 0.4 $\leq$ $\phi$ $<$ 0.5 is the reference 
spin phase. 
The plots from the top are the spin folded light curve (included for clarity),
  the variation in the hard X-ray normalisation, the column density of the 
partial covering absorber and its covering fraction.}
\label{fig:phspct}
\end{minipage}
\end{figure}

The resulting best fit parameter values from this analysis are shown
in Table \ref{tab:phspct}. The variations in the normalisation of the
optically thin plasma, the column density and partial covering
fraction are plotted in Figure \ref{fig:phspct}, together with the
folded light curve for the 0.5--10.0keV energy band of the {\sl ASCA}
data.  It can be seen that the column density does not vary
significantly over the whole spin cycle. The normalisation
parameter of the hard X-ray component shows a maximum over 0.2 $\leq$
$\phi$ $<$ 0.4, decreasing to a minimum at $\phi$ = 0.7. The covering
fraction of the partial absorber component shows a steady decline from
its maximum value at $\phi$ = 0.0 to a minimum value at $\phi$ = 0.5.
This pattern and phasing closely follows that of the {\sl RXTE}
8.0--25.0keV/2.0--8.0keV hardness ratio, shown in Figure 3.

Relating this back to the main features of the modulation of the light
curve shows that the maximum countrate (0.1 $\leq$ $\phi$ $<$ 0.4) is
due to a combination of increasing normalisation coupled with a
decreasing level of absorption. It would appear that an increase in
the normalisation factor is probably the main cause of the second
maximum (0.75 $\leq$ $\phi$ $<$ 0.85) of the light curve, but a
contribution from a decrease in the covering fraction may also play a
part.  The secondary minimum (0.55 $\leq$ $\phi$ $<$ 0.75) is mainly
due to a lower normalisation factor.  Finally, ``the dip'' can be
accounted for by the maximum value of the covering fraction, although,
again, a contribution from a slight decrease in the normalisation
factor cannot be ruled out.

\begin{table*}	
\begin{center}\small
\begin{tabular}{p{15mm}p{15mm}p{15mm}p{15mm}p{15mm}p{15mm}p{20mm}p{18mm}}
Spin Phase& N$_{H}(2)$             &Covering              & BB                   & MEKAL                & Gaussian:            & Flux(observed)    &$\chi^2_\nu$ (dof) \\
          & 10$^{22}$ cm$^{-2}$  &Fraction              & Norm:                & Norm:                & Norm:                & 2--10keV          &                 \\
          &                      &                      & 10$^{-4}$            & 10$^{-2}$            &10$^{-5}$             & 10$^{-11}$erg     &                 \\
          &                      &                      &                      &                      &                      & cm$^{-2}$ s$^{-1}$&                 \\
\hline
 0.0-0.1  &7.4$_{-1.5}^{+1.5}$   &0.80$_{-0.02}^{+0.02}$&0.48$_{-0.48}^{+1.03}$&1.88$_{-0.18}^{+0.21}$&7.96$_{-6.33}^{+2.94}$& 1.86              & 1.09 (149) \\
          &                      &                      &                      &                      &                      &                   &              \\ 
 0.1-0.2  &8.6$_{-2.2}^{+2.4}$   &0.70$_{-0.04}^{+0.03}$&1.62$_{-1.60}^{+0.57}$&2.10$_{-0.22}^{+0.26}$&11.9$_{-5.64}^{+3.20}$& 2.17              & 1.00 (198)   \\
          &                      &                      &                      &                      &                      &                   &              \\ 
 0.2-0.3  &5.4$_{-1.3}^{+2.5}$   &0.59$_{-0.03}^{+0.03}$&1.24$_{-0.54}^{+0.51}$&2.22$_{-0.16}^{+0.20}$&12.7$_{-5.69}^{+3.20}$& 2.59              & 1.11 (248)   \\
          &                      &                      &                      &                      &                      &                   &              \\ 
 0.3-0.4  &5.7$_{-1.5}^{+2.2}$   &0.55$_{-0.04}^{+0.03}$&1.63$_{-0.64}^{+0.45}$&2.21$_{-0.17}^{+0.21}$&7.77$_{-4.21}^{+4.53}$& 2.55              & 0.91 (245)   \\
          &                      &                      &                      &                      &                      &                   &              \\ 
 0.4-0.5  &8.1$_{-2.7}^{+3.9}$   &0.47$_{-0.05}^{+0.06}$&1.16$_{-0.38}^{+0.55}$&2.18$_{-0.20}^{+0.28}$&8.00$_{-5.09}^{+3.50}$& 2.49              & 1.12 (253)   \\
          &                      &                      &                      &                      &                      &                   &                \\ 
 0.5-0.6  &3.8$_{-1.5}^{+2.0}$   &0.43$_{+0.04}^{+0.04}$&1.47$_{-0.43}^{+0.55}$&1.79$_{-0.14}^{+0.17}$&7.02$_{-5.94}^{+2.68}$& 2.23              & 1.08 (241)   \\
          &                      &                      &                      &                      &                      &                   &              \\ 
 0.6-0.7  &6.9$_{-2.6}^{+4.1}$   &0.54$_{-0.06}^{+0.05}$&0.58$_{-0.35}^{+0.48}$&1.69$_{-0.17}^{+0.23}$&11.7$_{-5.42}^{+3.30}$& 1.96              & 0.98 (202)   \\
          &                      &                      &                      &                      &                      &                   &              \\ 
 0.7-0.8  &4.0$_{-1.3}^{+1.9}$   &0.52$_{-0.04}^{+0.03}$&0.65$_{-0.36}^{+0.49}$&1.61$_{-0.12}^{+0.13}$&6.69$_{-3.89}^{+3.51}$& 1.96              & 0.99 (209)   \\
          &                      &                      &                      &                      &                      &                   &              \\ 
 0.8-0.9  &8.1$_{-2.7}^{+3.9}$   &0.50$_{-0.05}^{+0.05}$&0.52$_{-0.46}^{+0.47}$&1.99$_{-0.18}^{+0.20}$&4.21$_{-4.06}^{+3.98}$& 2.23              & 1.04 (231)   \\
          &                      &                      &                      &                      &                      &                   &              \\ 
 0.9-0.0  &4.8$_{-1.1}^{+1.5}$   &0.58$_{-0.03}^{+0.03}$&1.42$_{-0.55}^{+0.48}$&1.87$_{-0.12}^{+0.16}$&11.8$_{-5.24}^{+3.20}$& 2.2               & 0.95 (218)   \\
\hline
\end{tabular}
\caption{The variable parameter values from spin phase-resolved spectral 
analysis over 10 phase-bins. The reference model (0.4 $\leq$ $\phi$ $<$ 0.5 ) 
was
 simultaneously fitted to the medium and high bit rate data from both of the 
SIS instruments and high bit rate data from the GIS2 instrument on the  
{\sl ASCA} satellite.The normalisation and  absorption parameters only were 
varied from those of the reference model. 
The blackbody, bremssstralung normalisations and absorption include the error 
range at the 90\% confidence level. The  blackbody, Mekal and Gaussian 
normalisations are standard for the relevant {\it xspec} models. Also included
 are  the  $\chi^2_\nu$ and observed flux (i.e. not corrected for absorption) 
for the 10 phases.  } 
\label{tab:phspct}
\end{center}
\end{table*}

\section{Mass of the white dwarf}
\label{mass}

\subsection{Procedure}

In recent years much work has been done in determining the shock
temperature in magnetic CVs (e.g. Cropper, Ramsay \& Wu 1998, Fuijmoto
\& Ishia 1996). One of the main goals of this work has been to
estimate the mass of the white dwarf since these two parameters are
closely linked.  To estimate the mass of the white dwarf in PQ Gem we
use the emission model of Cropper et al. (1999) (CWRK) to fit data
from the RXTE PCA detector (since it has an extended high energy range
- the temperature of the hard X-ray component was poorly constrained
using the ASCA data because of the 10keV upper limit). This model was
developed to fit high quality X-ray spectra appropriate to the
conditions found in the accretion sites of MCVs. It is a more complex
model than the generalised optical thin plasma code we used for the
phase-resolved spectroscopy.

The parameter, $\epsilon_{s}$ (the ratio of the cyclotron to bremsstralung 
cooling at the shock), was fixed at 0.5 which, from figure 1b of 
Wu, Chanmugam \& Shaviv (1995) corresponds to a magnetic field strength of 
$\sim15$MG (Piirola et al. 1993, V\"{a}th et al. 1996, Potter et al. 1997). We 
stratified the volume between the surface of the white dwarf and 
the shock front (Cropper et al. 1998) into 100 levels. 
The free parameters during the spectral fitting were the specific accretion
rate in g cm$^{-2}$s$^{-1}$, and the mass of the white dwarf. A
viewing angle is also required to scale the reflection component in the
model. The abundance was allowed to vary from a solar profile in the final
stages of the fitting.
 
\subsection{The Results}

The {\sl RXTE} spectral fitting was restricted to the spin phase range
of the data to 0.1 $\leq$ $\phi$ $<$ 0.5 which corresponds to that
portion of the spin cycle where a single emission site is visible
(Potter et al. 1997). This phase range also excludes the dip centered
on $\phi=$0.0 which from {\it ROSAT} and {\it ASCA} data, has been
shown to be due to absorption. Models including a homogeneous absorber 
components together with a CWRK emission component  were found to be
sufficient to produce an acceptable fit ($\chi^2_{\nu}$ $\sim$0.89) to
this data using an estimated average viewing angle of
$\sim$60$\degr$. This gave a white dwarf mass of 1.21M$_{\sun}$ (Table
\ref{tab:massfit}). We discuss this further in \S \ref{dmass}.
              
\begin{table*}	
\begin{center}\small
%\begin{minipage}[t]{90mm}
\begin{tabular}{p{10mm}p{10mm}p{10mm}p{20mm}p{12mm}p{12mm}p{21mm}p{15mm}}
Satellite &detector &angle   &N$_{H,cold}(1)$     & $\dot{m}$         &rc         & WD mass          & $\chi^2_\nu$ (dof)\\
          &         &$\degr$ &10$^{22}$ cm$^{-2}$ & gcm$^{-2}$s$^{-1}$&10$^{7}$cm & $M_{\sun}$       &       \\
\hline
{\sl RXTE}&PCA      &$\sim$60& 5.2                & 2.3               &9.5        & 1.21 (1.16-1.28) & 0.89 (46)  \\
\hline
\end{tabular}
\caption{Results from fitting the CWRK model to the {\sl RXTE} PCA spectrum 
(0.1 $\leq$ $\phi$ $<$ 0.5).Details of the spectral fitting process are given 
in \S \ref{mass} }  
\label{tab:massfit}
\end{center}
\end{table*}

\section{Discussion}

In section~\ref{phasespect} we find that the spin phase modulation can
be modelled well by a variation in the partial covering fraction of a
neutral absorber at low energies and in the normalisation at higher
energies. Before we discuss this further, we briefly examine whether
the modulation at both high and low energies can both be explained by
a tall accretion shock.

In the case where the shock has significant vertical extent, the lower
(cooler) part of the shock could be obscured by the limb of the white
dwarf at certain spin phases.  The light curves will then have a
larger amplitude at low energies than at high energies (e.g. Allan,
Hellier \& Beardmore 1998). The height of the accretion column, $H$, can
be estimated from the relationship \begin{displaymath} H=5.45\times
10^8\dot{M}^{-1}_{16}f_{-2}M_{WD}^{3/2}R_{WD}^{1/2} \end{displaymath}
(Frank, King \& Raine 1991). Using the lower limit of the fractional
area and the corresponding $\dot{M}$ derived in \S \ref{doccult}
(cf Table \ref{tab:mdot}) gives an upper bound of $H$=5.4$\times$10$^7$cm
or 0.14 R$_{WD}$. However, the observational evidence suggests that PQ
Gem has a significant magnetic field (\S \ref{intro}) which implies
that the shock height will be lower than this (e.g. Cropper et
al. 1999) due to cyclotron cooling. This suggests that this mechanism is not 
the cause of the variation. However, the accretion regions are
likely to be sufficiently structured (e.g. Potter et al 1997) so that
such a scenario cannot be excluded. On the other hand, because the
spectral variations can be well modelled by a variation in covering
fraction and normalisation, we go on to consider this explanation in
more detail.

\begin{table*}	
\begin{center}\small
%\begin{minipage}[t]{90mm}
\begin{tabular}{p{18mm}p{21mm}p{13mm}p{26mm}p{13mm}p{13mm}p{15mm}}
$\dot{m}$            & WD mass        &$\chi^2_\nu$ (dof)&unabsorbed flux          & luminosity        & $\dot{M}$         & fractional   \\
                     &                &                  &(0.001-100.0keV)         &                   &                   & area          \\
(g cm$^{-2}$s$^{-1}$)& $(M_{\sun})$   &                  &(ergs cm$^{-2}$ s$^{-1}$)& (ergs s$^{-1}$)   & (g s$^{-1}$)      &               \\
\hline
0.5                  & 1.20(1.13-1.26)&0.91(52)          &6.3$\times 10^{-11}$     &1.2$\times 10^{32}$&3.0$\times 10^{15}$&3.1$\times 10^{-3}$\\
1.0                  & 1.21(1.16-1.24)&0.89(52)          &6.4$\times 10^{-11}$     &1.2$\times 10^{33}$&2.9$\times 10^{15}$&1.6$\times 10^{-3}$\\
5.0                  & 1.22(1.15-1.29)&0.86(52)          &6.6$\times 10^{-11}$     &1.3$\times 10^{33}$&2.8$\times 10^{15}$&3.4$\times 10^{-4}$\\
\end{tabular}
\caption{The effect fixing $\dot{m}$ to a range of values in the CWRK model during the spectral fitting. Details are given in 
\S \ref{doccult}.}  
\label{tab:mdot}
\end{center}
\end{table*}

\subsection{Spin Pulse Modulation at Higher Energies}
\label{dspinmod_h}

It is evident from an inspection of the 8--25 keV spin-phased {\sl RXTE} light
curve (Figure 2) and the spin-phased normalisation in the {\sl ASCA}
spectroscopy (Figure 5) that they are broadly similar. As this emission is
expected to be at most weakly beamed, this indicates that the variation in
intensity is caused by changes in visibility of the X-ray emitting region as
the white dwarf rotates. The phase of maximum emission therefore corresponds to
the phase of maximum visibility of the emitting region. There is possibly a
slight phase shift between the two curves, with a clear peak at $\phi$ = 0.2
in the {\sl RXTE} light curve, and a more extended maximum around 
0.2 $\leq$ $\phi$ $<$ 0.4 in the {\sl ASCA} normalisation, but it is unclear 
given the uncertainties in the normalisations whether this is significant.
Spin phases 0.2--0.4 also correspond to the phases at which the accretion
region in the upper hemisphere is seen most close to face on (Potter et al
1997, figure 9). This suggests that this is principle cause of the spin pulse 
modulation at higher energies.

\subsection{Spin Pulse Modulation at Lower Energies}
\label{dspinmod_l}

At lower energies, the maximum in the covering fraction of the
absorber at $\phi$ = 0.0 is the major effect on the soft X-ray light
curve, while the local increase at $\phi$ = 0.6 combined with the
decrease in normalisation at this phase to cause the second minimum
(\S \ref{phasespect}). Potter et al. (1997) find that accretion occurs
preferentially along field lines which thread the disc ahead of the
accreting pole (their figure 12). Material accreting along these field
lines can therefore be identified as the source of the absorption at
phase 0.0. Similarly, at $\phi$ = 0.6, in the absence of absorption by
the disc (cf \S \ref{doccult}), the local increase in covering
fraction is likely to be caused by absorbing material along field
lines which intersect the line of sight to the lower pole (Potter et
al. 1997, figure 9).

Figure \ref{fig:phspct} indicates that the change in absorption can be 
explained {\it entirely} by
a change in covering fraction, with the column density of the absorber
remaining constant. With
reference to figure 9 of Potter et al. (1997), at $\phi$ = 0.0, all of the
field lines between the disc and the accreting pole will intersect the line of
sight. At $\phi$ = 0.2, those field lines to the leading part of the accretion
region will no longer intersect the line of sight, while by $\phi$ = 0.4,
only field lines to the trailing part of the accretion region will do so. If
the column density through accreting field lines is $\sim 5\times 10^{22}$ 
cm$^2$, the observed variation in covering fraction in Figure \ref{fig:phspct}
 can be reproduced.

The accretion flow may be more finely fragmented along field lines
feeding the leading edge of the accretion region. This would be
expected because finely fragmented material is threaded by the
magnetic field more easily than the larger denser inhomogeneities
(e.g. Wickramasinghe 1988).  From considerations of the packing
fraction, the line of sight through more finely fragmented material is
less likely to pass between gaps in the flow than in the case of the
larger inhomogeneities. This effect would  reproduce the high
covering fraction at $\phi$ = 0.0, and its subsequent decrease towards
$\phi$ = 0.5.

\subsection{Accretion Model}

The accretion scenario we are proposing fits neither the standard
occultation model of King \& Shaviv (1984) nor the accretion curtain
model of Rosen, Mason \& C\'{o}rdova (1988). The orientation of the
accretion region at spin phase maximum is as predicted by the
occultation model, but with absorption effects modifying the light curve
significantly. In this it is similar to the ``weak field/fast
rotator'' model of Norton et al. (1999) with the symmetry of the
accretion curtain about the magnetic axis modified by the leading
field lines preferentially stripping material from the inner margin of
the disc. The high (among IPs) magnetic field in PQ Gem and relatively
high inclination and low dipole offset ($\sim 60^{\circ}$ and $\sim
30^{\circ}$, Potter et al. 1997) ensure that material on accreting
field lines travels far enough out of the orbital plane to pass
through the line of sight to the accretion region, causing the
observed absorption effects. In this it differs substantially from EX
Hya which possesses a magnetic field $<$ 1MG. 

\subsection{Occultation by the Accretion Disc?}
\label{doccult}

Finally, we check whether the accretion disk can extend close enough
to the white dwarf for it to have an affect on the X-ray light curves
by obscuring the lower emission region.

Using the Ghosh \& Lamb formulation (Li, Wickramasinghe \& Rudiger 1996) the
radius to the truncated inner edge of the accretion disc, $r_A$, is given by
\begin{eqnarray*}
r_A = 0.52\mu_{WD}^{4/7}(2GM)^{-1/7}\dot{M}^{-2/7} 
\end{eqnarray*}
where $\mu_{WD}$ is the magnetic moment of the white dwarf. 

The magnetic moment can be estimated from the relationship $B =
\mu/r^3$. Using the fits to the {\sl RXTE} data and the model of CWRK
we found a best fit to the white dwarf radius,
$R_{WD}$, of $3.8\times 10^8$ cm.  Hence, with a magnetic field
strength, $B_{WD}$, of 15MG (Piirola et al. 1993, V\"{a}th et
al. 1996, Potter et al. 1997), $\mu_{WD} = 9.6\times10^{32}$ G
cm$^{3}$. The accretion luminosity,
\begin{eqnarray*}
L_{acc} = GM\dot{M}/R_{WD}
\end{eqnarray*}
where $M$ is the mass of the white dwarf and L$_{acc}$ is emitted mostly 
 in the X-ray energy band, enables estimation of the accretion rate, $\dot{M}$.
 
The unabsorbed spectral model from the analysis of the integrated
spectrum (\S \ref{intspct}) extrapolated for the energy range
0.001--100.0keV gives the X-ray flux at $3.1\times10^{-10}$erg
cm$^{-2}$ s$^{-1}$ which, taking the distance to PQ Gem of 400pc
(Patterson 1994), gives a luminosity of 6.0$\times10^{33}$erg s$^{-1}$
and hence a mass transfer rate rate, $\dot{M}$, of $1.4\times10^{16}$g
s$^{-1}$. This is typical for IPs (Warner 1995).  The resulting $r_A =
1.3\times 10^{10}$ cm or $\sim34R_{WD}$ may be too large given that an 
estimate of the distance to the first Lagrangian point is $\approx$ 200 
R$_{WD}$ (Plavec \& Kratochvil 1964) (the main uncertainty in $r_A$
is in the magnetic moment $\mu_{WD}$). Nevertheless it does indicate
that with a system inclination of $60\degr$ (Potter et al. 1997), the
line of sight to the white dwarf surface is likely to be clear of the
accretion disc at all spin phases, and this is unlikely to contribute to the 
cause of the covering fraction variation.

\subsection{The size of the Accretion Region}

  Using the radius of the white dwarf,
the specific mass accretion rate, $\dot{m}$, (2.3 g cm$^{-2}$s$^{-1}$
cf Table \ref{tab:massfit}) and $\dot{M}$ determined above we can
derive a fractional accretion area, $f$, of 9.0$\times 10^{-3}$.  This
is within the normal expectation of 0.001$\la f \la$ 0.02 for an IP
(Rosen 1992). However, from the {\sl RXTE} data $\dot{M}$ =2.9 $\times
10^{15}$ g s$^{-1}$ which implies a fractional area of only 7.1
$\times 10^{-4}$. Refitting the {\sl RXTE} data to the model in which
$\dot{m}$ was fixed at 0.5, 1.0 and 5.0 g cm$^2$ did not give a
significant adverse effect on the fit ($\chi^2_\nu$=0.91, 0.89, 0.86,
respectively). Table \ref{tab:mdot} gives the implied fractional area
and $\dot{M}$ for these $\dot{m}$ as well as the results from the
spectral fitting. For a low specific mass accretion rate we find that
the implied fractional area is consistent with the lower limit
determined using previous observations.

\subsection{The Mass of the White Dwarf}
\label{dmass}

Previous determinations of the mass of the white dwarf in PQ Gem using
an emission model fitted to {\it Ginga} data (Cropper et al. 1998,
1999) gave estimates $\geq$ 1.1 $M_{\sun}$. In the case of the IP XY
Ari, Ramsay et al. (1998) found that there was a good correspondence
between the estimates given by this model and those from eclipse
mapping.  In our work it is found that the {\sl RXTE} data gives
estimates which are very much better constrained than those made with
the {\sl ASCA} SIS data.  The estimate from our {\sl RXTE} data of
$M_{WD}$ = 1.21 (1.16-1.28) $M_{\sun}$ corresponds well to that of
1.21 ($>1.08) M_{\sun}$ obtained with {\it Ginga} data. Although this
appears to be unusually high Ramsay (2000) found that the white dwarf
in magnetic CVs were biased towards higher masses compared to isolated
white dwarfs.

\section*{Acknowledgements}

We gratefully acknowledge Darragh O'Donoghue for the use of his period
analysis software and also the referee, Chris Done, for her helpful comments.

 \end{document}